\documentstyle[epsf,psfig,epic,eepic,amssymb]{article}
\def\:{\hbox{\bf :}}

\def\thesection{\arabic{section}}
\newcommand{\sect}[1]{\stepcounter{section}\section*{\protect{
\noindent\large\bf\thesection~\normalsize\bf #1}}}

\def\thefigure{\arabic{figure}}
\def\fnum@figure{{\bf Figure \thefigure}}

\vspace*{1mm}
\begin{document}
\begin{figure}[hbt]
\begin{center}\epsfxsize=.7 \hsize\leavevmode\epsffile{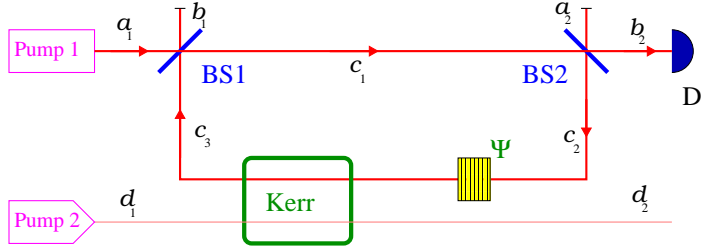}
\end{center}
\caption{\scriptsize Sketch of the experimental setup for the
generation of optical Fock states in the traveling wave mode $d_2$,
starting from two coherent pumps. The scheme is based on a conditional
{\sf ON-OFF} photodetector D, a non linear Cross--Kerr medium and a
phase shifter $\psi$. The cavity input radiation modes are labelled
$a$, the output modes $b$, the cavity modes $c$ and the signal mode
$d$. The quantum efficiency of the photodetector is not a crucial
parameter, as it only affects the production rate of the device and
not the quality of the generated states.}
\label{f:scheme}\end{figure}

\begin{center}{\Large\bf State preparation by photon filtering}\end{center}
\begin{center}by {\large{\sl  G. M. D'Ariano, L. Maccone, M. G. A. Paris, 
and M. F. Sacchi, }}\end{center}
\begin{center}
Theoretical Quantum Optics Group -- INFM Unit\'a 
di Pavia \\ Dipartimento di Fisica 'Alessandro Volta' -- Universit\'a di 
Pavia \\ via A. Bassi 6,  I-27100 Pavia, Italy\\
\end{center}
\begin{quote}{\bf Abstract.}
We propose a setup capable of generating Fock states of a single mode
radiation field. The scheme is based on coupling the signal field to a
ring cavity through cross-Kerr phase modulation, and on conditional
{\sf ON-OFF} photodetection at the output cavity mode.  The same setup
allows to prepare selected superpositions of Fock states and entangled
two-mode states. Remarkably, the detector's quantum efficiency does
not affect the reliability of the state synthesis.\end{quote}
%%%%%%%%%%%%%%%%%%%%%%%%%%%%%%%%%%%%%%%%%%%%%%%%%%%%%%%%%%%%%%%%%%%%%%%%%%%%% 
\sect{Introduction} 
The generation of optical radiation number (Fock) states is of
importance for a number of different applications, ranging from high
resolution spectroscopy to fundamental tests of Quantum Mechanics. In
quantum communication, Fock states achieve the optimal capacity of
quantum channels \cite{qcomm}, whereas in optical quantum computation
superpositions of Fock states are needed as input states
\cite{qcomput}.  The synthesis of Fock states also allows to
experimentally characterize active optical devices \cite{ham}.  In
this paper, we suggest a novel preparation scheme for Fock states and
their superpositions. In our proposal, a traveling wave mode is
coupled to a ring cavity through cross-Kerr phase modulation. The
cavity mode serves as a probe, and a successful photodetection at the
cavity output reduces the signal mode to a predetermined output
state. The proposed setup can also be used to synthesize two-mode
entangled states by using an additional Kerr medium, with possible
applications in quantum information and quantum teleportation
technology.  The scheme works with high-Q cavities and conventional
{\sf ON-OFF} photodetectors, whereas it needs relatively large
nonlinearities. These should be available in the next-future
technology, since recent theoretical \cite{imamoglu} and experimental
\cite{nature} developments indicate that huge Kerr phase shifts of
the order of $1$ radiant per photon can be obtained through
electromagnetic induced transparency.
%%%%%%%%%%%%%%%%%%%%%%%%%%%%%%%%%%%%%%%%%%%%%%%%%%%%%%%%%%%%%%%%%%%%%%%%%%%%% 
\sect{Experimental apparatus and Fock state synthesis} 
The scheme, depicted in Fig. \ref{f:scheme}, is based on a ring
cavity, coupled to the external radiation modes through two very low
transmissivity $\tau$ beam splitters (BS) and a cross Kerr-medium. A
tunable phase shifter, which shifts the field in the cavity by a phase
$\psi$, is included in the device. One of the two cavity input modes
($a_1$) is pumped with a coherent source, while the other one ($a_2$)
is left in the vacuum. Of the two output modes, one ($b_2$) is
monitored with an {\sf ON-OFF} photodetector D (which measures the
presence or absence of the radiation field), the other one ($b_1$) is
ignored, so that it is traced out. As will be shown in the following,
whatever is the state of the input signal mode, a number state will be
found at the output mode $d_2$ conditioned to a successful measurement
at the detector D. Upon fulfillment of certain conditions shown in the
following, superpositions of number states can also be generated. In
principle, the specific number state (or superposition) that will be
created is controlled by tuning the phase $\psi$.
\par
We now show how the device works by analyzing the dynamics of its
components in the Heisenberg picture. In principle, it would be
necessary to quantize the field in the cavity starting from its
classical equations of motion, as for example in
\cite{loudon}. However, we show that the same results are more easily
obtained with a simple one mode model of the cavity, by simply
identifying the output mode $c_2$ at the beam splitter BS2 with the
input mode $c_3$ at BS1 (see Fig. \ref{f:scheme}).  Let $a_{1,2}$
denote the two input modes for the ring cavity, $b_{1,2}$ the two
output modes, $c_1$ the cavity mode exiting BS1, $c_2$ the cavity mode
exiting BS2, $c_3$ the cavity mode exiting the Kerr medium and the
phase shifter, and $d_1$ and $d_2$ the input and output signal mode
respectively. In the Heisenberg picture, the input--output relations
which characterize the components of the scheme are
\begin{eqnarray}
&&\mbox{BS1:\ }
\left\{\matrix{c_1&=&\sqrt{\tau}a_1+\sqrt{1-\tau}c_3\cr 
b_1&=&-\sqrt{1-\tau}a_1+\sqrt{\tau}c_3
}\right.
\;,\label{iorelat}
\\ \nonumber&&
\mbox{BS2:\ }
\left\{\matrix{c_2&=&\sqrt{\tau}a_2+\sqrt{1-\tau}c_1\cr 
b_2&=&-\sqrt{1-\tau}a_2+\sqrt{\tau}c_1
}\right.
\;,
\\ \nonumber&&
\mbox{Kerr and Phase shift:\ }
c_3=c_2\; \exp\left[-i(\chi t d_1^\dagger d_1-\psi)\right]\;,
\end{eqnarray}
where $\tau$ is the transmissivity of the two beam splitters, $\chi t$
is the Kerr susceptibility multiplicated by the interaction time, and
$\psi$ is the phase shift imposed to the cavity field mode by the
phase shifter.  From Eqs. (\ref{iorelat}) the output modes $b_{1,2}$
of the cavity can be expressed as a function of the input cavity and
signal mode as \begin{eqnarray}
\left\{
\matrix{b_1&=&\kappa(\varphi)a_1+e^{i\varphi}\sigma(\varphi)a_2\cr 
b_2&=&\sigma(\varphi)a_1+\kappa(\varphi)a_2}\right.\;,\ \mbox{where\ }
\left\{\matrix{
\kappa(\varphi)&\doteq&\frac{\sqrt{1-\tau}(e^{i\varphi}-1)}
{1-(1-\tau)e^{i\varphi}}\cr
\sigma(\varphi)&\doteq&\frac{\tau}
{1-(1-\tau)e^{i\varphi}}\cr}\right.
\;\label{b1},
\end{eqnarray}
where $\varphi\doteq -\chi t d_1^\dagger d_1+\psi$ is the overall
phase shift due to the Kerr medium and the phase shifter. Notice that
the coupling with the signal mode is in the dependence of $\varphi$ on
the signal input mode $d_1$.
\par

From the Heisenberg evolution of the modes (\ref{b1}), one obtains the
state $|\Psi_{out}\rangle$ in the cavity output modes (before the
measurement) by using the creation operators $b_{1,2}^\dagger$ and by
expressing the input state $|\Psi_{in}\rangle$ on a Fock basis of the
input modes $a_{1,2}$:
\begin{eqnarray}
|\Psi_{in}\rangle=\sum_{n,m=0}^\infty\;c_{nm} 
\frac{(a_1^\dagger)^n(a_2^\dagger)^m}{\sqrt{n!m!}}
|0\rangle|0\rangle
%|n\rangle_{a_1}|m\rangle_{a_2} 
%=\sum_{n.m=0}^\infty\;c_{nm}
%\frac{(a_1^\dagger)^n(a_2^\dagger)^m}{\sqrt{n!m!}}|0\rangle|0\rangle
\ \longrightarrow\ |\Psi_{out}\rangle=\sum_{n,m=0}^\infty\;c_{nm}
\frac{(b_1^\dagger)^n(b_2^\dagger)^m}{\sqrt{n!m!}}|0\rangle|0\rangle
\;\label{evolut}.
\end{eqnarray}
Consider the overall input state $\varrho_{in}$ composed by a coherent
state in mode $a_1$, vacuum in $a_2$ and generic state in $d_1$ (which
will be described by the density matrix $\nu_{ss'}$ in the Fock
basis), {\it i.e.} \begin{eqnarray}
\varrho_{in}=|\alpha\rangle_{a_1}{}_{a_1}
\langle\alpha|\otimes|0\rangle_{a_2}{}_{a_2}
\langle 0|\otimes
\sum_{s,s'=0}^\infty\;\nu_{ss'}|s\rangle_{d_1}{}_{d_1}
\langle s'|
\;\label{psiin}.
\end{eqnarray}
Through Eq. (\ref{evolut}) we obtain the output state (before
detection by detector D) on the modes $b_{1,2}$ and $d_2$ as
\begin{eqnarray}
&&\varrho_{bd}=
|\alpha\kappa(\varphi)\rangle_{b_1}{}_{b_1}
\langle\alpha\kappa(\varphi)|\otimes
|\alpha e^{i\varphi}\sigma(\varphi)\rangle_{b_2}{}_{b_2}
\langle\alpha e^{i\varphi}\sigma(\varphi)|\otimes
\sum_{s,s'=0}^\infty\;\nu_{ss'}|s\rangle_{d_1}{}_{d_1}
\langle s'|=\nonumber \\ 
&&\sum_{s,s'=0}^\infty\;\nu_{ss'}
|\alpha\kappa(\varphi_s)\rangle_{b_1}{}_{b_1}
\langle\alpha\kappa(\varphi_{s'})|\otimes
|\alpha e^{i\varphi}\sigma(\varphi_s)\rangle_{b_2}{}_{b_2}
\langle\alpha e^{i\varphi}\sigma(\varphi_{s'})|\otimes
|s\rangle_{d_2}{}_{d_2}\langle s'|
\label{psiout},
\end{eqnarray}
where $\varphi_n\doteq -\chi t n+\psi$. After tracing out mode $b_1$,
the state $\varrho_{bd}$ becomes
\begin{eqnarray}
&\varrho'_{bd}=\label{outstate}\sum_{ss'}&\nu_{ss'}
e^{-\frac{|\alpha|^2}
2\left[|\kappa(\varphi_s)|^2+|\kappa(\varphi_{s'})|^2-
2\kappa(\varphi_s)\bar\kappa(\varphi_{s'})\right]}\times\\ \nonumber&&
|\sigma(\varphi_s)\alpha e^{i\varphi}\rangle_{b_2}{}_{b_2}
\langle\sigma(\varphi_{s'})\alpha e^{i\varphi}|\otimes
|s\rangle_{d_2}{}_{d_2}
\langle s'|\;,
\;
\end{eqnarray}
The state (\ref{outstate}) now undergoes a reduction at detector D. An
{\sf ON-OFF} photodetector, with quantum efficiency $\eta$ is described by
the two value probability operator measure (POM) \begin{eqnarray}
\Pi_{\mbox{\tiny \sf OFF}}=
\sum_{k=0}^\infty(1-\eta)^k|k\rangle\langle k|\;;\ 
\Pi_{\mbox{\tiny \sf ON}}=1-\Pi_{\mbox{\tiny \sf OFF}}
\;,\label{pom}
\end{eqnarray}
which is obtained from the Mandel-Kelley-Kleiner formula \cite{mkk}.
The final state after the reduction in mode $b_2$, in the case of a
successful ({\sf ON}) photodetection is given by
\begin{eqnarray}
&&\varrho_{out}=\frac{
\mbox{Tr}_{b_2}[\Pi_{\mbox{\tiny \sf ON}}\varrho_{bd}]}
{\mbox{Tr}[\Pi_{\mbox{\tiny \sf
ON}}\varrho_{bd}]}=\label{measurement}\\ \nonumber&&\frac
{e^{-|\alpha|^2}}{\cal N} \sum_{s,s'=0}^\infty\;\nu_{ss'}
\exp\left[{|\alpha|^2\left(\kappa(\varphi_s)\kappa^*(\varphi_{s'})+
\sigma(\varphi_s)\sigma^*(\varphi_{s'})e^{i(\varphi_s-\varphi_{s'})}
\right)}\right]\times\\ \nonumber &&\ \ \ \ \ \ \ \
\left(1-e^{-\eta|\alpha|^2\sigma(\varphi_s)\sigma^*(\varphi_{s'})}\right)
|s\rangle\langle s'|
\;,
\end{eqnarray}
where $\cal N$ is a normalization constant. Notice that the phase
factors in the exponential in (\ref{measurement}) do not play any
role, as, in the working regime we exploit, these factors tend to one.
\par
\begin{figure}[hbt]
\begin{center}\epsfxsize=.65 \hsize\leavevmode\epsffile{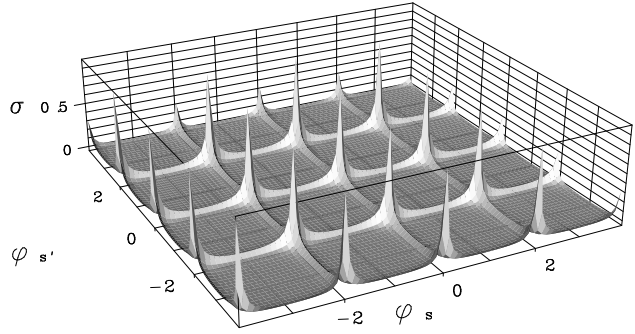}
\end{center}
\caption{\scriptsize Plot of the function
$|\sigma(\varphi_s)\sigma^*(\varphi_s')|$ [$\varphi_s$,
$\varphi_{s'}$ in $\pi$ units.], with $\tau=.1$,
$n^*=\frac\psi{\chi t}=0$ and $l^*=\frac{2\pi}{\chi t}=2$. Notice the
``fakir chair'' structure: the function approaches a sum of delta
functions for beam splitter transmissivity $\tau\to 0$.}
\label{f:3dsigma}\end{figure}

\noindent In the limit $\tau\to 0$, the effective transmissivity
$\sigma(\varphi)$ of the cavity tends to a sum of Kronecker
deltas\begin{eqnarray}
\lim_{\tau\to 0}|\sigma(\varphi_s)|=
%\lim_{\tau\to 0}\frac\tau{\sqrt{(1-\tau)^2-2(1-\tau)\cos\varphi_s+1}}=
\left\{\matrix{&1&\ \mbox{for}\ \varphi_s\doteq -\chi t s+\psi=2k\pi\ \ (k\in{\mathbb Z})\cr
&0&\ \mbox{for}\ \varphi_s\neq2k\pi}\right.= 
\sum_{k=-\infty}^\infty\;
\delta_{s,kl^*+n^*} 
\;\label{sigma},
\end{eqnarray}
where $l^*=\frac{2\pi}{\chi t}$ is the distance between the peaks, and
$n^*=\frac\psi{\chi t}$ is the position of the first peak. For the values
of $l^*$ and $n^*$ for which the quantity $kl^*+n^*$ is not an
integer, there is no contribution to the sum in (\ref{sigma}). In
fig. \ref{f:3dsigma} the function
$|\sigma(\varphi_s)\sigma^*(\varphi_s')|$ is plotted {\it vs.}
$\varphi_s$ and $\varphi_s'$.

\begin{figure}[hbt]
\begin{center}\epsfxsize=.3 \hsize\leavevmode\epsffile{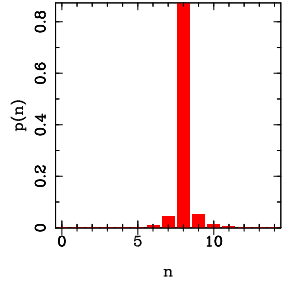}
\end{center}
\caption{\scriptsize Plot of photon number distribution of a
synthesized Fock state. The experimental parameters are: $\tau=0.01$,
$\psi=.4$, $\chi t=0.05$, $\eta=10\%$ and the two pumps (in modes $a_1$
and $d_1$) are in a coherent state with amplitude
$\alpha=\beta=3$. The output state approaches a Fock state
$|8\rangle$.}
\label{f:fockprod}\end{figure}

From Eqs. (\ref{measurement}) and (\ref{sigma}) it is clear that for
small values of the Kerr susceptibility ${\chi t}$ ({\it i.e.} $l^*$ high
enough so that $\nu_{il^*+n^*,jl^*+n^*}\simeq 0$ for both $i$ and $j$
nonzero) and for high pump value $|\alpha|\gg 1$, the output state
approaches a Fock state:
\begin{eqnarray}
\lim_{\tau\to 0}\varrho_{out}=\;|n^*\rangle\langle n^*|
\;\label{fockst}.
\end{eqnarray}
Notice that the state $|n^*\rangle$ to be synthesized can be
controlled by varying the phase $\psi$. In Fig. \ref{f:fockprod} the
photon number distribution of the state (\ref{measurement}), is
given. Notice how the limiting case of Eq. (\ref{fockst}) is
approached.
\par
\begin{figure}[hbt]
\begin{center}\epsfxsize=.3 \hsize\leavevmode\epsffile{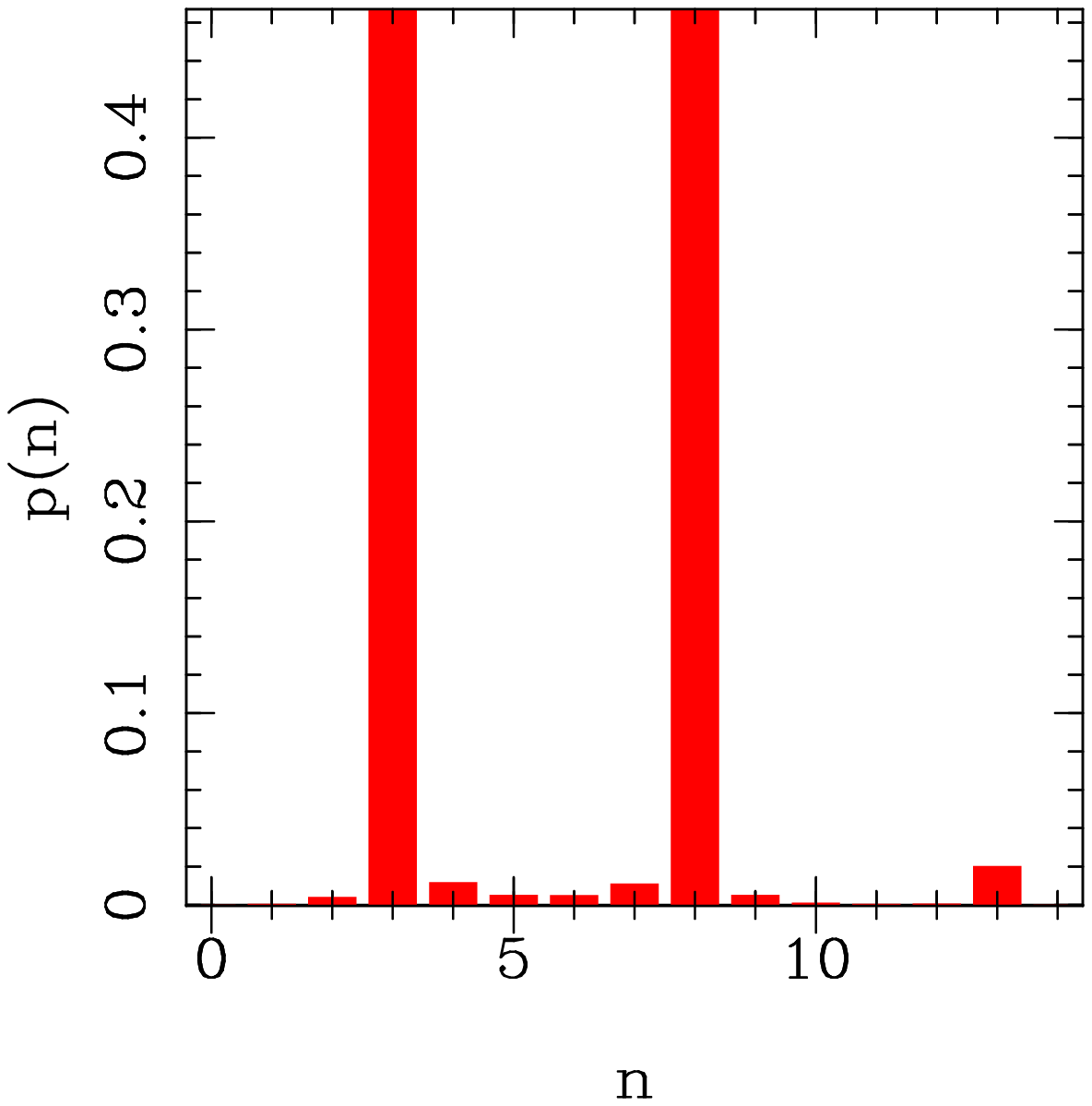}
\epsfxsize=.3 \hsize\leavevmode\epsffile{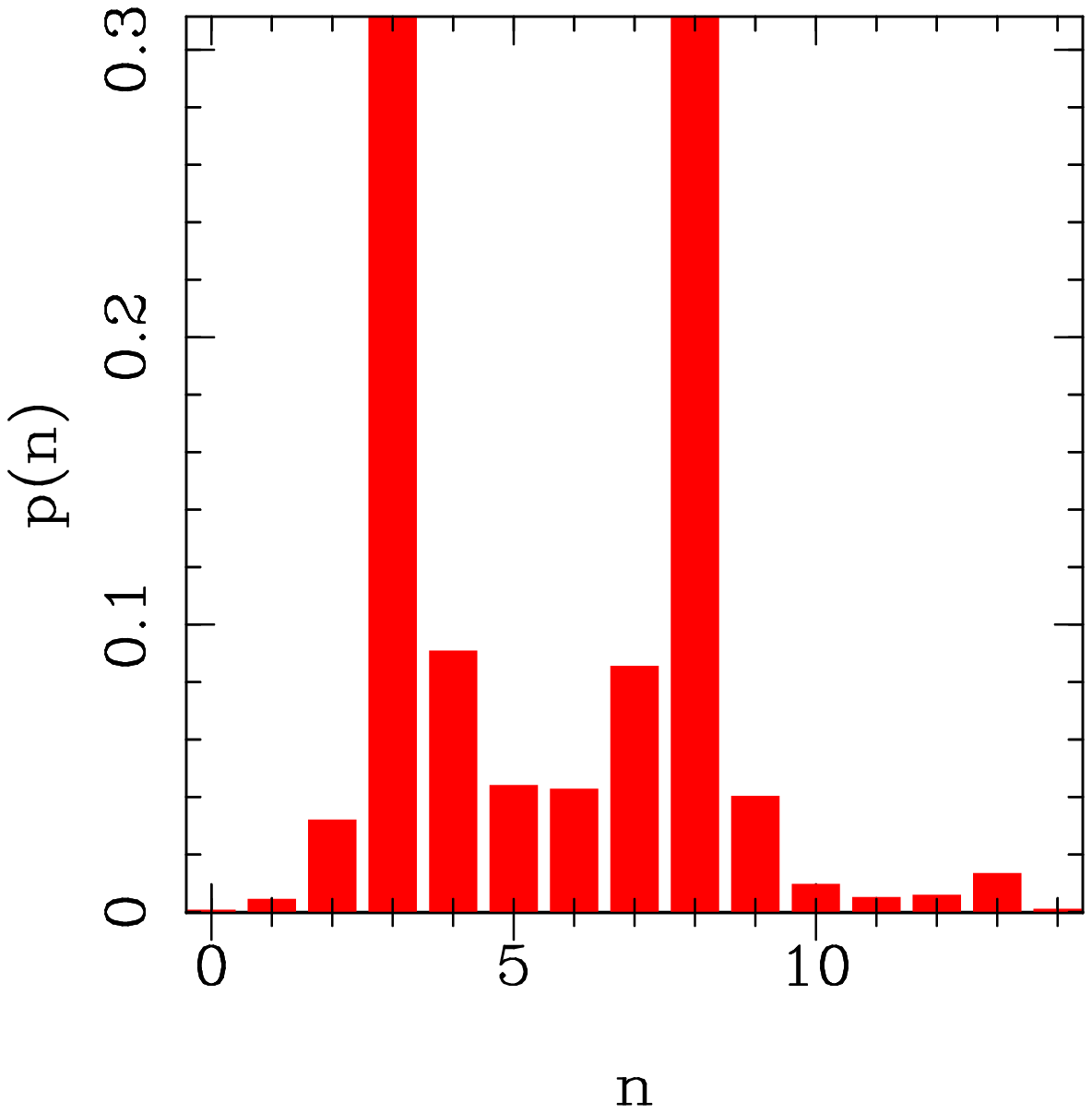}
\end{center}
\caption{\scriptsize Plot of the diagonal of the density matrix
$\varrho_{out}$ exiting the device for a coherent input in mode $d_1$
with modulus $|\beta|^2=\root^5\of{\frac{8!}{3!}}$ and with $l^*=5$,
$n^*=3$. The parameters are chosen so to have the superposition $\frac
1{\sqrt{2}}(|3\rangle+|8\rangle)$. In the left plot, notice a residual
component at the Fock state 13, which results from the term $k=2$ in
the sum (\ref{sigma}). Here, the quantum efficiency of detector D is
$\eta=10\%$, the modulus of the cavity pump is $\alpha=8$, the BS
transmissivity is $\tau=.06$. In the right plot we used $\tau=.2$, to
show the smearing effect of a bad cavity.}
\label{f:superp}\end{figure}
\begin{figure}[hbt]
\begin{center}\epsfxsize=.65 \hsize\leavevmode\epsffile{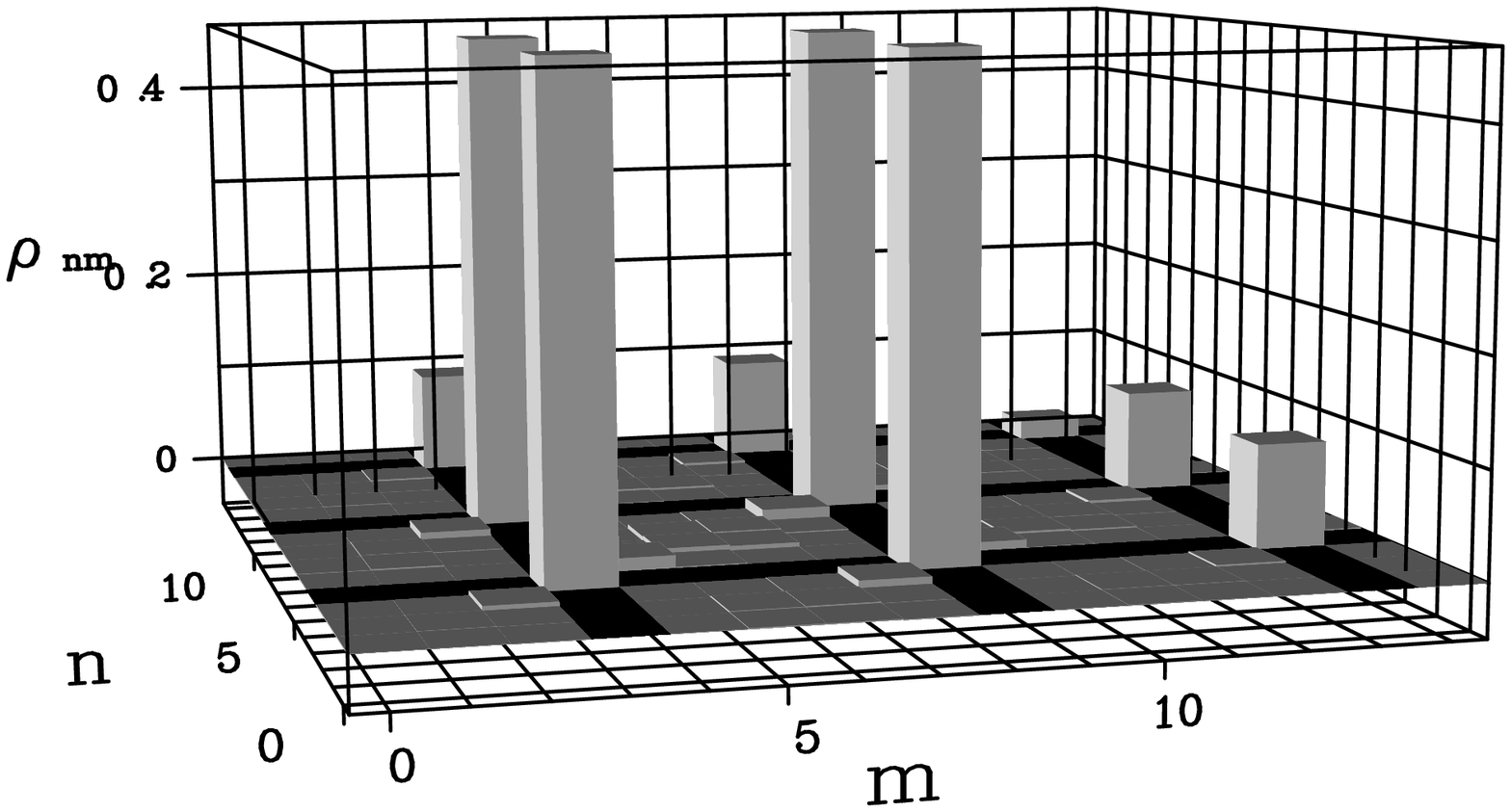}
\end{center}
\begin{center}\epsfxsize=.65 \hsize\leavevmode\epsffile{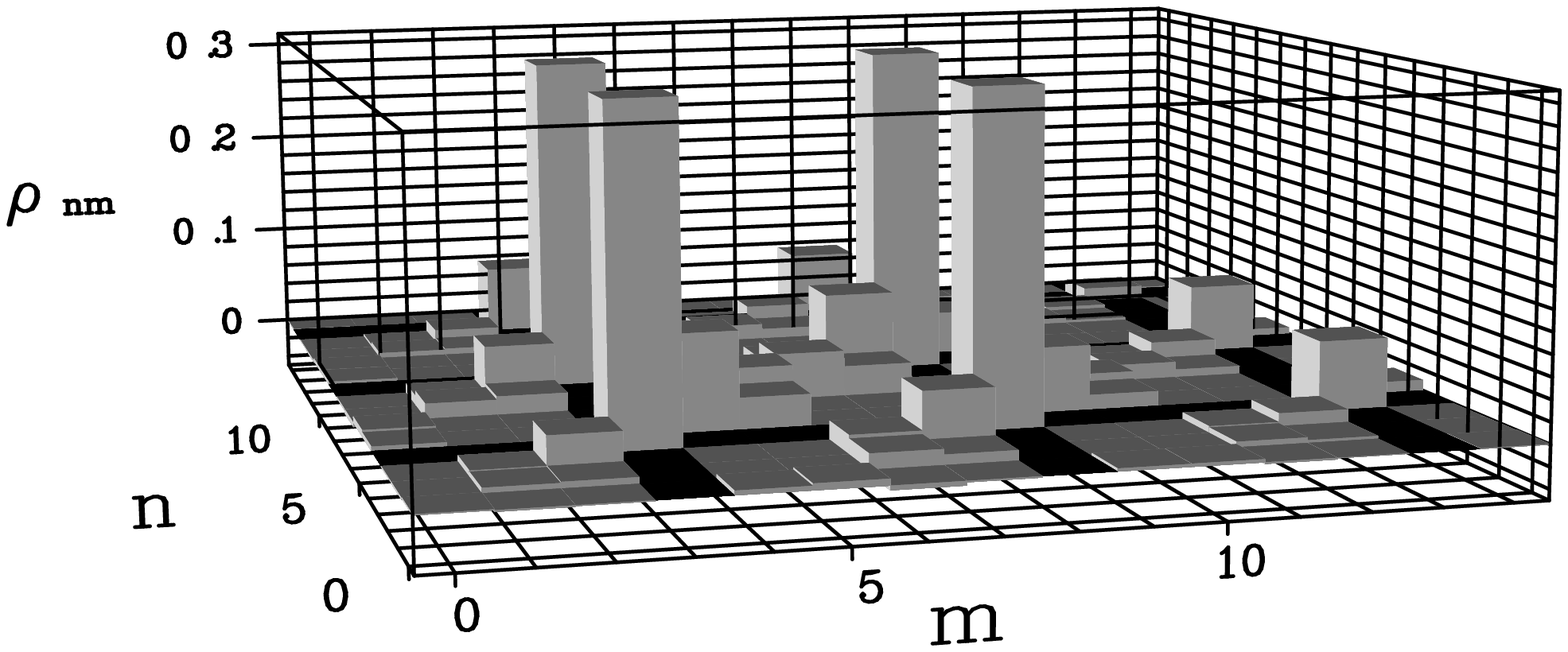}
\end{center}
\caption{\scriptsize Density matrix for the two states given in
Fig. \ref{f:superp}.}
\label{f:superp3d}\end{figure}

\sect{Other applications}
%SUPERPOSITIONS
%%%%%%%%%%%%%%%%%%%%%%%%%%%%%%%%%%%%%%%%%%%%%%%%%%%%%%%
{\bf Superposition generation.}  In order to produce superpositions of
selected equally spaced number states, one has to provide higher
values for the Kerr nonlinearity $\chi t$, or alternatively to provide
sufficiently excited input state in mode $d_1$, so that the input
state coefficients $\nu_{il^*+n^*,jl^*+n^*}$ are different from zero
for different values of $i$ and $j$.  For example, as shown in
Figs. \ref{f:superp} and
\ref{f:superp3d}, starting from a coherent input state with density 
matrix $\nu_{ss'}=e^{-|\beta|^2}
\frac{\beta^s(\beta^*)^{s'}}{\sqrt{s!s'!}}$ it is possible to generate 
the superposition \begin{eqnarray}
|\Psi\rangle=\frac 1{\sqrt{2}}(|n^*\rangle+e^{i\Phi}|n^*+l^*\rangle)
\;\label{superp}
\end{eqnarray}
by choosing $|\beta|^2=\root^{l^*}\of{\frac{(n^*-l^*)!}{n^*!}}$ and
$\arg \beta=\frac\Phi{l^*}$.
It must be stressed that only superpositions of the type \begin{eqnarray}
|\Psi\rangle=\sum_{k=0}^\infty c_k|n^*+kl^*\rangle
\;\label{allowedsuperp}
\end{eqnarray}
may be created, where the coefficients $c_k$ are determined by the
state incoming in mode $d_1$, {\it i.e.} by the coefficients
$\nu_{ss'}$.

\vskip 1\baselineskip
\par

%DISTRIBUTION AND STATE MEASUREMENT
%%%%%%%%%%%%%%%%%%%%%%%%%%%%%%%%%%%%%%%%%%%%%%%%%%%%%%%
{\bf Photon number and state measurement.} The proposed device can
also be used for the measurement of the photon number distribution of
an unknown quantum state. The probability of successful ({\sf ON})
detection, in fact, is given by \begin{eqnarray}
P_{\mbox{\tiny\sf ON}}=\mbox{Tr}[\Pi_{\mbox{\tiny\sf ON}}\rho_{bd}]=\sum_{k=0}^\infty\nu_{kk}(1-
e^{-\eta|\alpha|^2|\sigma(\varphi_s)|^2})
\;\label{probsuccess},
\end{eqnarray}
and for sufficiently high $l^*$ and $|\alpha|\gg 1$, is proportional
to one of the diagonal elements of the input density matrix
$P_{\mbox{\tiny \sf ON}}\simeq\nu_{n^* n^*}$. Thus, if an ensemble of
identical states is impinged into the device in mode $d_1$, by tuning
the cavity to different values of $n^*$ and measuring the relative
frequency of {\sf ON} photodetection, one can obtain
$\nu_{n^*n^*}$. This allows the measurement of the photon number
distribution of an arbitrary state, using less experimental data and
lower quantum efficiency detectors than in homodyne tomography (which
is the currently used experimental procedure for reconstructing the
number distribution of arbitrary states). Moreover, by using the
least-squares inversion method outlined in \cite{opatr} and by
displacing the input state with a beam splitter, it is possible to
reconstruct also the truncated density matrix of the state in the Fock
basis. The details for such an experimental setup will be given
elsewhere.
\vskip 1\baselineskip
\par

\begin{figure}[hbt]
\begin{center}
\epsfxsize=.7\hsize\leavevmode\epsffile{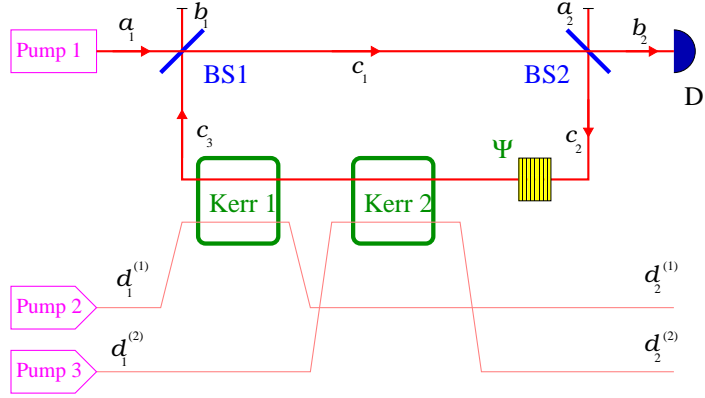}
\end{center}
\caption{\scriptsize Modification of the proposed experimental setup,
in order to construct entangled states between the output modes
$d_2^{(1)}$ and $d_2^{(2)}$. }
\label{f:modif}\end{figure}

%TWO-MODE ENTANGLED STATES.
%%%%%%%%%%%%%%%%%%%%%%%%%%%%%%%%%%%%%%%%%%%%%%%%%%%%%%%
{\bf Entanglement creation.} A modified version of the proposed experiment may also be used for the 
production of two-mode entangled states. One only has to use two Kerr
crystals inside the cavity, as shown in Fig. \ref{f:modif}.
In this case, the total phase shift imposed by the cavity is given by
the sum of the effect of each Kerr medium. For identical Kerr crystals
$K1$ and $K2$ ({\it i.e.} $\chi_{K1}\equiv\chi_{K2}\doteq\chi$), one
has $l^*=\frac{\pi}{\chi t}$ and $n^*=\frac\psi{2\chi t}=n_1^*+n_2^*$,
where $n_1^*$ and $n_2^*$ are the eigenvalues of the number states in
the modes $d^{(1)}$ and $d^{(2)}$ impinging into each of the Kerr
crystals. In this case, Eq.  (\ref{sigma}) becomes
\begin{eqnarray}
\lim_{\tau\to 0}|\sigma(\varphi_s)|=\sum_k\delta_{s,kl^*+n_1^*+n_2^*}\;
\label{sigmaentang},
\end{eqnarray}
and the output state (\ref{fockst}) for $\tau \ll 1$ and $|\alpha|\gg
1$ now reads
\begin{eqnarray}
\varrho_{out}=\sum_{n,k=0}^{n^*}
\nu^{(1)}_{nk}\nu^{(2)}_{nk}|n\rangle_{d^{(1)}_2}{}_{d^{(1)}_2}
\langle k| \otimes
|n^*-n\rangle_{d^{(2)}_2}{}_{d^{(2)}_2}\langle n^*-k|
\;\label{fockstent},
\end{eqnarray}
that is obtained in the limit of low Kerr nonlinearity $\chi t\ll 1$
({\it i.e.} $l^*\gg 1$). For input signal pure states
$|\Psi_{in}\rangle=\sum_k\nu^{(1)}_k|k\rangle_{d_1^{(1)}}\otimes
\sum_j\nu^{(2)}_j|j\rangle_{d_1^{(2)}}$,
one finds \begin{eqnarray}
|\Psi_{out}\rangle=\sum_{k=0}^{n^*}\nu^{(1)}_k\nu^{(2)}_{n^*-k}
|k\rangle_{d_2^{(1)}}|n^*-k\rangle_{d_2^{(2)}}\;\label{fockstent2}.
\end{eqnarray}
It is obvious that any multipartite entanglement between many modes
could be synthesized in principle, by increasing the number of Kerr
media in the cavity. This method could then be used also to generate
Greenberger--Horn--Zeilinger states.
\par
\begin{figure}[hbt]
\begin{center}
\epsfxsize=.3 \hsize\leavevmode\epsffile{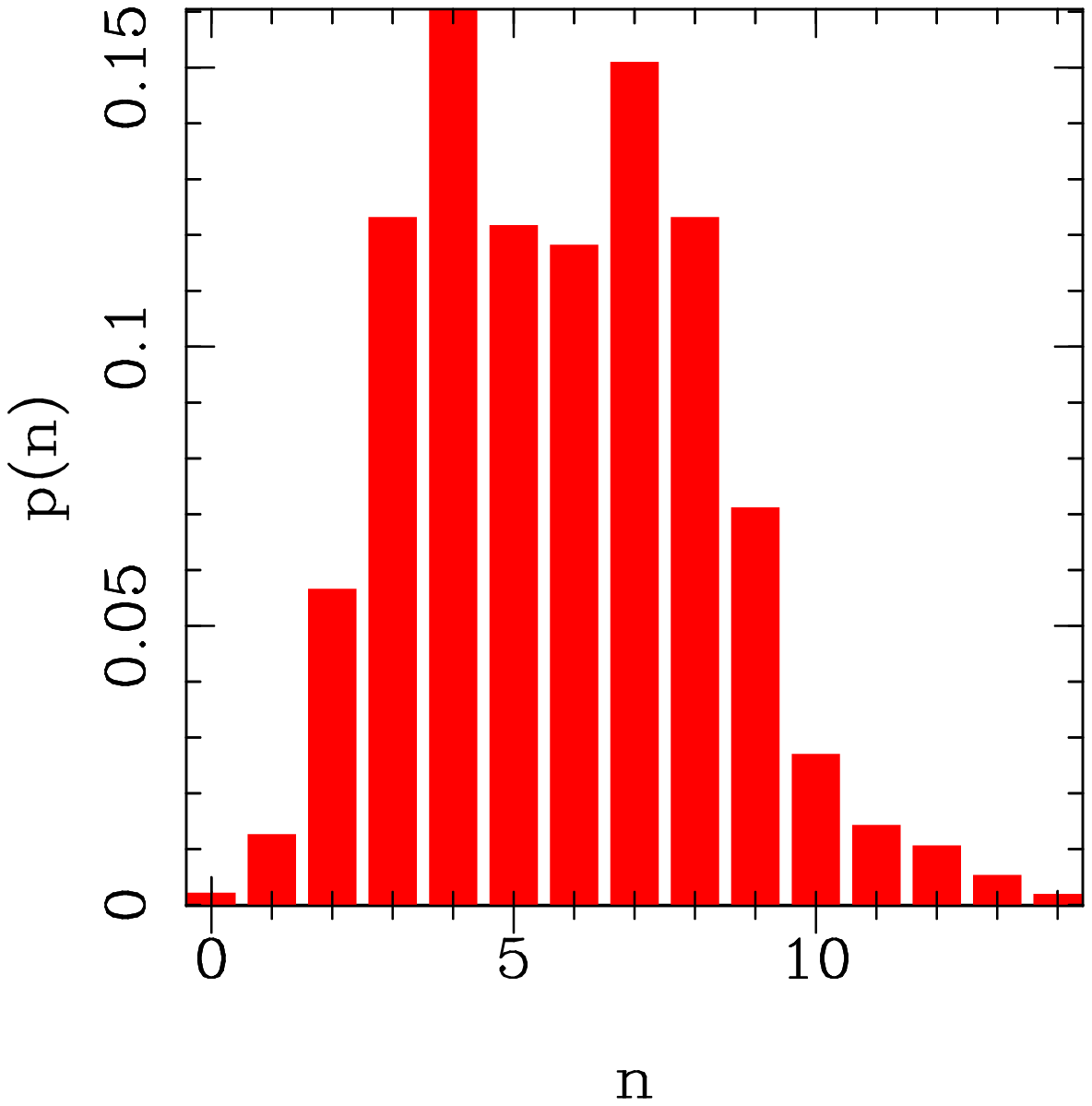}
\epsfxsize=.3 \hsize\leavevmode\epsffile{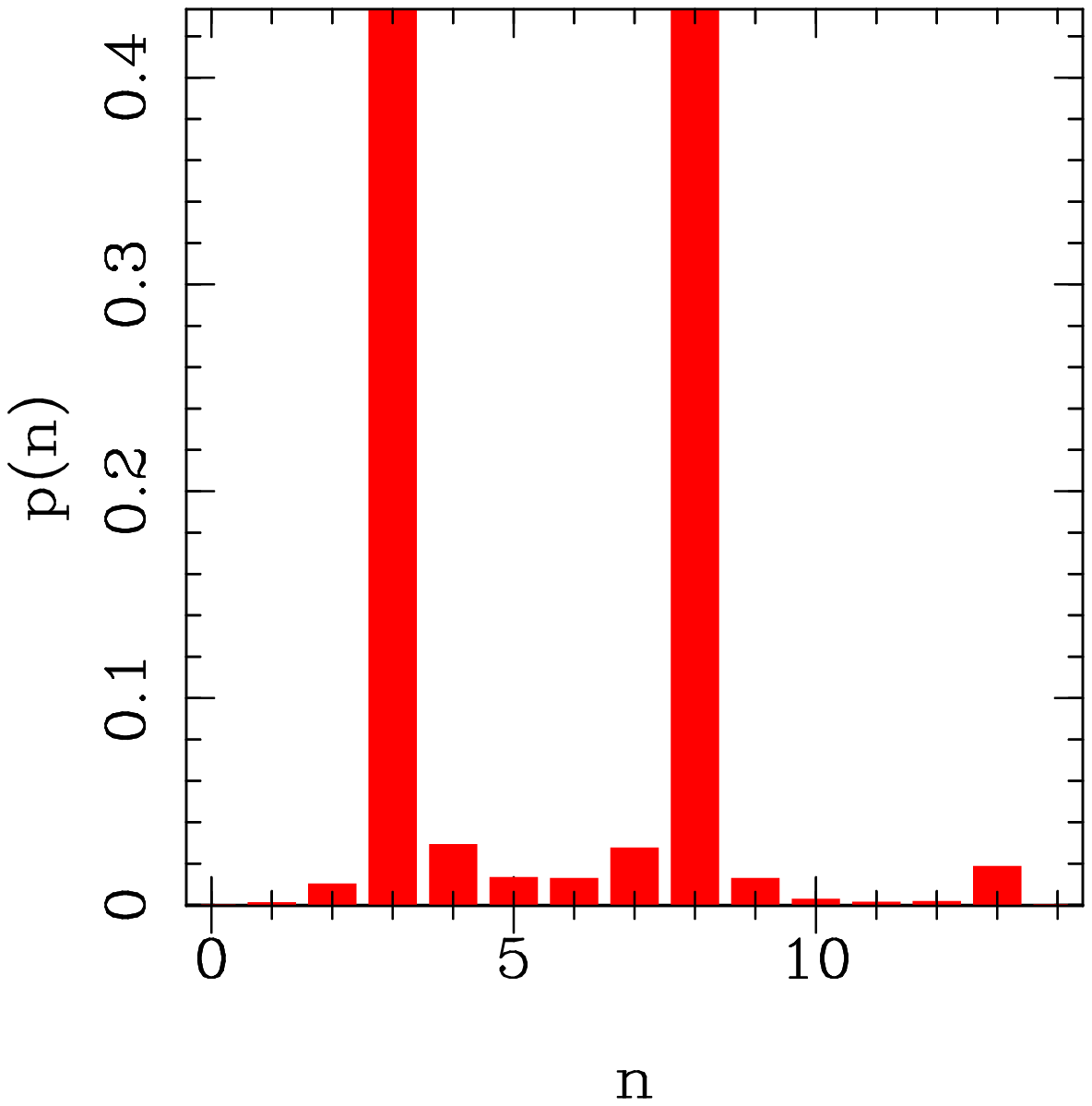}
\end{center}
\caption{\scriptsize Plots for the same parameters as for the right 
figure \ref{f:superp} ({\it i.e.}
$|\beta|^2=\root^5\of{\frac{8!}{3!}}$, $l^*=5$, $n^*=3$, $\alpha=8$
and $\tau=.2$). On the left $\eta=100\%$, while on the right
$\eta=1\%$. This shows how lowering the detector's quantum efficiency
enhances the quality of the state. The cost is paid in terms of a
lower production rate: the probability (\ref{probsuccess}) of
obtaining the state depicted on the left is $P_{\mbox{\tiny\sf
ON}}\simeq 0.789$, while on the right $P_{\mbox{\tiny\sf ON}}\simeq
0.106$. }
\label{f:qeff}\end{figure}

\sect{Feasibility}
%QUANTUM EFFICIENCY
%%%%%%%%%%%%%%%%%%%%%%%%%%%%%%%%%%%%%%%%%%%%%%%%%%%%%%%
We now analyze the effect of detector's D quantum efficiency. As can
be seen from Fig. \ref{f:qeff}, lowering the detector's quantum
efficiency actually purifies the state. Qualitatively this can be
explained by noticing that in Eq. (\ref{measurement}) the lowering of
$\eta$ contributes to the convergence exploited in Eq. (\ref{fockst})
to produce the Fock state (or the superpositions). It would seem that
one could actually increase the output signal quality by using low
efficiency photodetectors. The drawback is given in terms of
production rate of the desired Fock states. It is obvious that it is
less probable for a non--efficient photodetector to click, so the
production rate is drastically decreased. In fact, the probability of
a successful {\sf ON} photodetection is given by
Eq. (\ref{probsuccess}), from which one can see that
$P_{\mbox{\tiny\sf ON}}$ exponentially decreases with decreasing
$\eta$, and which for $\tau\to 0$ and $|\alpha|\gg 1$ tends to
\begin{eqnarray} P_{\mbox{\tiny\sf ON}}\longrightarrow
\sum_{k=0}^\infty\nu_{kl^*+n^*,kl^*+n^*}
\;\label{tend}.
\end{eqnarray}
The quantum efficiency $\eta$ of the detector in the proposed device
is, therefore, not a critical parameter, so that ordinary
photodetectors may be used in the experimental setup. Using low
efficiency photodetection does not reduce the quality of the output
states, but only it affects their production rate.

%EXPERIMENTAL FEASIBILITY
%%%%%%%%%%%%%%%%%%%%%%%%%%%%%%%%%%%%%%%%%%%%%%%%%%%%%%%
The actual feasibility of the proposed experimental setup relies on
the availability of good ring cavities, suitable Kerr nonlinearities,
and ordinary photodetection. The cavity couples to the outside modes
through small transmissivity beam splitters $\tau\simeq0.1\div
0.01\%$. The Kerr nonlinearities that are needed are of the order of
$\chi t\simeq 0.01$ for the creation of number states and of the order
of $\chi t\simeq 0.1$ for obtaining
superpositions. Photodetection with quantum efficiency as low as
$\eta=1\%$ has been shown to be effective.  The output state must be
controlled by varying the phase shift $\psi$ which can be varied, in
ordinary experimental setups, in steps of the order of $\frac\pi{500}$
\cite{fase}. To our knowledge, the experiment for the generation of
Fock states or two mode entangled states should be feasible with
laboratory technology now available. The creation of Fock state
superposition may ask for Kerr nonlinearities which are not yet
available in the optical domain, though recent results
\cite{imamoglu,nature} indicate that giant non-linear shifts of the
order of $1$ radiant per photon may be obtained through
electromagnetically induced transparency.

\sect{Conclusions}
In conclusion we have proposed an optical device capable of creating
optical Fock states, selected superposition of Fock states, and two
mode entangled states. The experimental setup is composed of a high-Q
ring cavity coupled with the signal mode through a cross-Kerr medium
and an {\sf ON-OFF} photodetection. A successful photodetection
reduces the signal mode to a predetermined output state.  We have
shown that imperfect photodetection does not affect the quality of the
output states. The same device can be also used for the measurement of
the photon number distribution and of the density matrix in the Fock
basis.  The applications for such a device in the modern ``Quantum
technology'' are numerous and span through many of the fields of
advanced quantum optics research.

%\begin{thebibliography}{99}
\section*{\normalsize\bf References}
\begin{description}
\bibitem[1]{qcomm} H. P. Yuen and M. Ozawa, Phys. Rev. Lett. {\bf 70}, 
363 (1993).
\bibitem[2]{qcomput} I. L. Chuang and Y. Yamamoto, Phys. Rev. A {\bf
52}, 3489 (1995).
\bibitem[3]{ham} G. M. D'Ariano, L. Maccone, Phys. Rev. Lett. 
{\bf 80}, 5465 (1998).
\bibitem[4]{imamoglu}  H. Schmidt and A. Imamoglu, Opt. Lett. {\bf
21}, 1936 (1996); H. Schmidt and A. Imamoglu, Opt. Lett. {\bf 23},
1007 (1998).
\bibitem[5]{nature} L. V. Hau, S. E. Harris, Z. Dutton, and
C. H. Behroozi, Nature {\bf 397}, 594 (1999).
%\bibitem[6]{oneatomlas} K. An, J. J. Childs, R. R. Dasari, and
%M. S. Feld, Phys. Rev. Lett. {\bf 73}, 3375 (1994).
%\bibitem[4]{walther} D. Meschede, H. Walther, and G. Muller,
%Phys. Rev. Lett. {\bf 54}, 551 (1985).
\bibitem[6]{loudon} M. Ley and R. Loudon, J. Mod. Opt. {\bf 34}, 227
(1987).
\bibitem[7]{mkk} P. L. Kelley and W. H. Kleiner, Phys. Rev. A {\bf
30}, 844 (1964).
\bibitem[8]{fase} O. Haderka, M. Hendrych, private communication.
\bibitem[9]{opatr} T. Opatrn\`y and D. - G. Welsch, Phys. Rev. A {\bf
55}, 1462 (1997).
\end{description}
%\end{thebibliography}
\end{document}